\documentclass[a4paper]{article}
\usepackage{a4wide}
\usepackage[latin1]{inputenc}
\usepackage{amsmath}
\usepackage{amsfonts}
\usepackage{amssymb}
\usepackage{lmodern,dsfont}
\usepackage{graphicx,epsfig}
\usepackage{authblk}

\title{Connecting active and passive microrheology in living cells}

\author[1]{A. Dechant}
\author[1,2]{E. Lutz}
\affil[1]{Dahlem Center for Complex Quantum Systems, Freie Universit{\" a}t Berlin, 14195 Berlin, Germany}
\affil[2]{Institute for Theoretical Physics, University of Erlangen-N{\"u}rnberg, 91058 Erlangen, Germany}
\date{}

\begin{document}

\maketitle

\begin{abstract}
We use a model based on the fractional Langevin equation with external noise to describe the anomalous dynamics observed in  microrheology experiments in  living cells. 
This model reproduces both the subdiffusive short-time and the superdiffusive long-time behavior. 
We show that the former reflects the equilibrium properties of the cell, while the latter is due to the nonequilibrium external noise. 
This allows to infer the transport properties of the system under active measurements from the transient behavior obtained from passive measurements, extending the connection between active and passive microrheology to the nonequilibrium regime.
The active and passive results can be linked via a generalized Stokes-Einstein relation based on an effective time-dependent temperature, which can be determined from the transient passive behavior.
In order to reproduce experimental data, we further find that the external noise describing the active components of the cell has to be nonstationary. 
We establish that the latter leads to a time-dependent noise spectral density.
\end{abstract}

\bigskip

Particle tracking microrheology is a powerful tool to examine biological cells in a non-destructive manner \cite{gar05,wir09,squ10,kol11}. 
Small tracer particles are injected into the cytoplasm and useful information  about the mechanical properties and the internal dynamics of the cell can be gained  by observing their motion; important examples include elastic and viscous moduli \cite{mas95,mas97,sch97,che03,bra07} and  the characteristics of fluctuating forces \cite{lau03,wil08,gal09}. 
Two complementary approaches are usually  distinguished: In passive microrheology, the tracer particles are allowed to move freely through the cell, thus highlighting the diffusive features of the cytoplasm. 
On the other hand, in active microrheology, an external force is applied to the  particles and their response is recorded, revealing the cytoplasm's transport properties. 
For equilibrium systems, the diffusive and transport aspects are linked via the Stokes-Einstein relation (a consequence of the fluctuation-dissipation relation \cite{rei65}). 
Accordingly, the result of active measurements can be successfully deduced from the much simpler passive ones, and vice versa \cite{mas95,mas97,sch97,che03}. 
However, the Stokes-Einstein relation breaks down in nonequilibrium systems such as living cells \cite{lau03,wil08,gal09,bur05}, where the active nature of the cytoplasm plays an essential role \cite{squ10,miz07,miz08}. 
A striking example of this nonequilibrium nature are the observed diffusive properties of the cytoplasm: while tracer particles in a passive crowded equilibrium system  exhibit subdiffusion \cite{wei04,gol06}, active forces in  cells lead to a superdiffusive spreading \cite{gal09,bur05}. 
In view of the above, it did not seem possible to obtain transport properties of living cells from single-particle passive experiments \cite{lau03,wil08,gal09,bur05}.

In this paper, we develop a formalism that enables  to connect  passive and active single-particle microrheology in nonequilibrium systems, where the Stokes-Einstein relation does not hold, and apply it to  biological cells. To this end, we use a fractional Langevin equation, with an external noise term that models active forces inside the cell  \cite{bra09,bru09}.
We derive a generalized Stokes-Einstein relation by defining an effective time-dependent temperature  relating the mobility to the diffusivity. 
Using the latter, we elucidate how the transport properties can be obtained from the comparison of  the transient and long-time behavior observed in a passive measurement, without the need for active measurements. 
In addition, we show that the external noise force has to be nonstationary in order to explain  experimental data, that is, its fluctuating properties are not invariant under  time shift  \cite{rei65}. 
We find that this nonstationarity  leads to a time-dependent noise spectral density which we compute explicitly. 
The latter should be observable experimentally, revealing a new facet of cell dynamics.

\section{Generalized Langevin equation \label{SEC2}}

The diffusive dynamics of a tracer particle inside living cells has been  successfully described by a generalized Langevin equation of the form \cite{lau03,wil08,gal09,bru09},
\begin{align}
m \ddot{x}(t) = - \int_{0}^{t} {\rm d}t' \gamma(t-t') \dot{x}(t') + F_{\rm p}(t) + \eta_{\rm i}(t) \label{GLE},
\end{align}
where $x(t)$ is the position of the particle, $m$ its mass, $\gamma(t)$ a retarded friction kernel describing the delayed back-action of the viscous medium, $F_{\rm p}(t)$ a probe force (in the active measurement scheme) and $\eta_{\rm i}(t)$  a  thermal noise force describing random collision events between the tracer and constituents of the environment---hence the subscript "i" for internal. 
At thermal equilibrium, the noise correlation function and the friction kernel are connected by the fluctuation-dissipation relation \cite{rei65},
\begin{align}
\langle \eta_{\rm i}(t) \eta_{\rm i}(t') \rangle = k_B T \gamma(t-t')  \label{FDT},
\end{align}
where $\langle \ldots \rangle$ denotes an ensemble average, $T$ the temperature and $k_B$  the Boltzmann constant. Equation \eqref{GLE} is called fractional Langevin equation, if the friction kernel is given by a power law \cite{man68,lut01},
\begin{align}
\gamma(t) = \frac{\gamma_{\alpha}}{\Gamma(\alpha)} t^{\alpha-1} \quad \text{with} \quad 0 < \alpha < 1 \label{friction},
\end{align}
where $\Gamma(\alpha)$ is the Gamma function and $\gamma_{\alpha}$  a generalized friction coefficient. 
The exponent $\alpha$ characterizes the dynamics of the system: For $\alpha \rightarrow 0$, we recover the usual memory-less Langevin equation corresponding to a purely viscous liquid, while for $\alpha \rightarrow 1$, Eq.~\eqref{GLE} describes an elastic system. 
For intermediate values of $\alpha$ between $0$ and $1$, the system is  thus referred to as viscoelastic. 
The internal noise in this case corresponds to fractional Gaussian noise, i.e. Gaussian noise with power law correlations.

The diffusive behavior observed in the passive measurement scheme may be  characterized by the mean-square displacement \cite{rei65},
\begin{align}
\langle \Delta x^2(t) \rangle = \langle (x(t)-x(0))^2 \rangle - \langle x(t)-x(0) \rangle^2 \label{MSDdef}.
\end{align}
On the other hand, the time-dependent viscoelastic response of a system to a  variation  of the probe force $F_{\rm p}(t)$ during an  active measurement  is given by  the creep function $J(t)$ defined as \cite{bur05,kol11},
\begin{align}
\langle x(t) - x(0) \rangle = \int_{0}^{t} {\rm d} t' J(t-t') \frac{{\rm d} F_{\rm p}(t')}{{\rm d} t'} + J(t) F_{\rm p}(0). \label{creepdef}
\end{align}
Using the fractional Langevin equation \eqref{GLE},  the long-time behavior, $t \gg (m/\gamma_{\alpha})^{1/(\alpha+1)}$, of the two quantities is found to be \cite{lut01,pot03},
\begin{align}
\langle \Delta x^2(t) \rangle %= \frac{2 k_B T}{m} t^2 E_{\alpha+1,3}\left(-\frac{\gamma_{\alpha}}{m} t^{\alpha+1} \right)% 
&\simeq \frac{2 k_B T}{\gamma_{\alpha} \Gamma(2-\alpha)} t^{1-\alpha} \label{MSDeq} , \\
J(t) %= \frac{1}{m} t^2 E_{\alpha+1,3}\left(-\frac{\gamma_{\alpha}}{m} t^{\alpha+1} \right) % 
&\simeq \frac{1}{\gamma_{\alpha} \Gamma(2-\alpha)} t^{1-\alpha} \label{creepeq} .
\end{align}
These two equations immediately lead to,
\begin{align}
\langle \Delta x^2(t) \rangle = 2 k_B T J(t) \label{einstein},
\end{align}
which is an expression of the Stokes-Einstein relation  linking diffusivity (the time derivative of the mean-square displacement) and mobility (the time derivative of the creep function) \cite{gar05,wir09,squ10,kol11}. 
The above equality is a fundamental property of equilibrium systems and, from a practical point of view, allows to infer the response of the system to an external perturbation solely by studying its diffusive behavior. 
Since $0 < \alpha < 1$, the equilibrium system is always subdiffusive.

\begin{figure}[t]
\begin{center}
\includegraphics[width=0.9\textwidth, trim=19mm 10mm 26mm 15mm, clip]{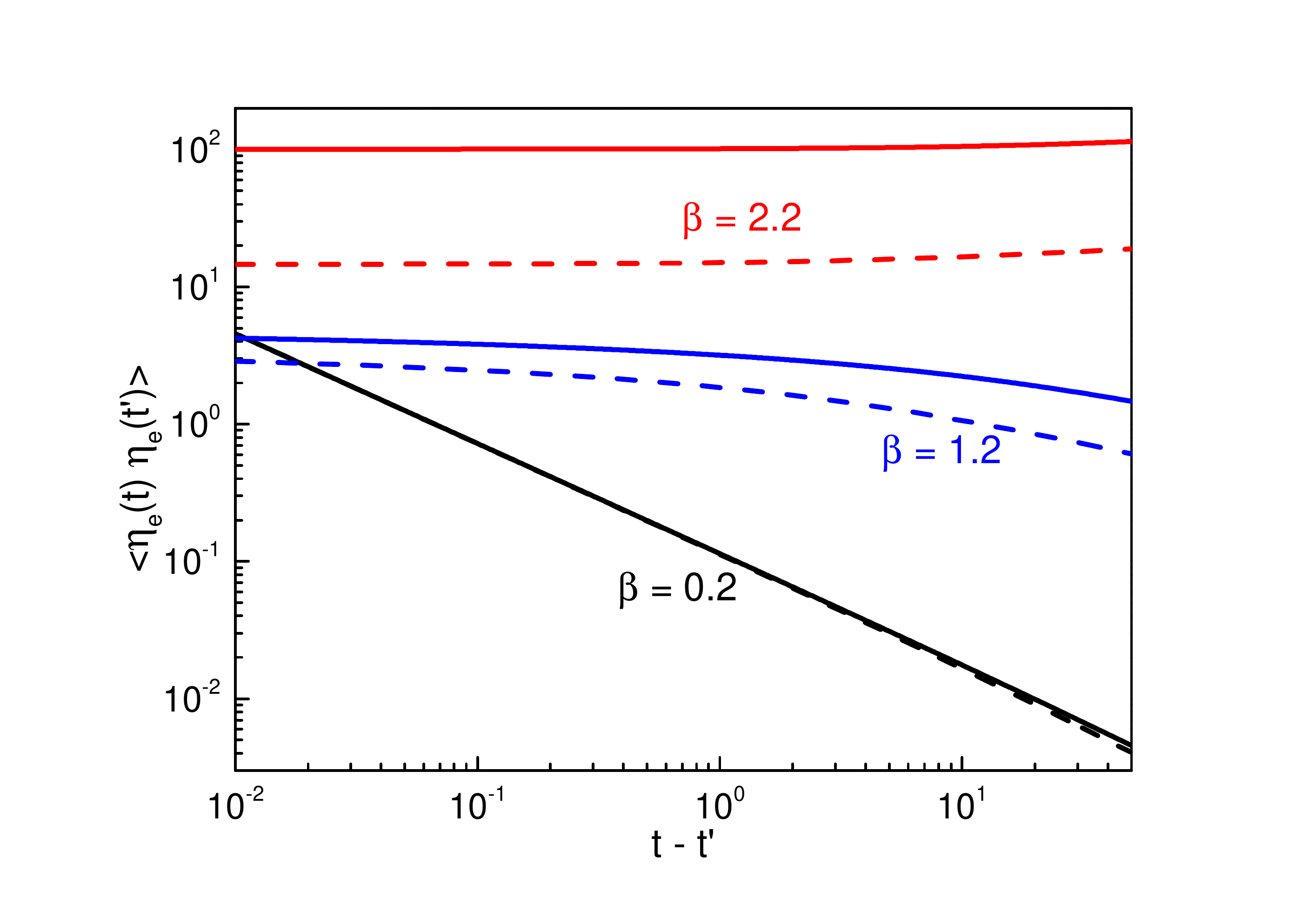}
\caption{Autocorrelation function of the external nonequilibrium noise \eqref{noisecorr} for different values of the exponent $\beta$ as a function of the time difference $t-t'$. The dashed and solid lines correspond to $t' = 10$ and $t' = 50$, respectively. For $\beta < 1$ (black), the noise is stationary and its correlation function depends only on the time difference in the long-time limit. For $1 < \beta < 2$ (blue), the variance of the noise increases with time $t'$ and its correlation function is nonstationary, i.e. explicitly dependent on $t'$. Finally, for $\beta > 2$ (red), the noise correlation function increases both as a function of the absolute time $t'$ and of the difference $t-t'$, indicating that the process is becoming more correlated with time. \label{fig0}}
\end{center}
\end{figure}

In living cells, however, a tracer particle is not only acted upon by the internal thermal noise, it also experiences forces due to the active components of the cytoplasm, e.g. molecular motors moving along microtubuli or actin filaments. While  these often lead to straight, directed motion, it has been realized recently that they also result in random, diffusive-like motion \cite{miz08,bra09,bru09}.
We model the latter by introducing an additional, external noise term $\eta_{\rm e}(t)$ into the Langevin equation \eqref{GLE} following Ref.~\cite{bru09}.
This external noise does not satisfy the fluctuation-dissipation relation \eqref{FDT} and drives the system out of equilibrium. The expression for the creep function \eqref{creepeq} will  be left unchanged by the presence of the unbiased external noise, but there will be an additional contribution to the mean-square displacement \eqref{MSDeq}. In order to reproduce the measured power-law behavior of the noise power spectrum \cite{lau03,wil08,gal09}, the correlation function of the external noise was  taken to be  of the form \cite{bru09},
\begin{align}
\langle &\eta_{\rm e}(t) \eta_{\rm e}(t') \rangle \sim (t-t')^{\beta-1} \label{noisecorr_old},
\end{align}
with $0 < \beta < 1$. The above expression was shown to quantitatively describe the experimentally observed transition between subdiffusion and superdiffusion of melanosomes  in Xenopus laevis melanocytes, as well as to provide a good estimate of in vivo motor forces \cite{bru09}. However, the correlation function  \eqref{noisecorr_old} is stationary, as it only depends on the time difference $t-t'$. We will show below that it  fails to account for the nonstationary behavior observed in a majority of experiments \cite{tre07,tre08}. In the following, we will hence use a more general form of the  noise correlation function which, in  Laplace space, reads, 
\begin{align}
\langle \tilde{\eta}_{\rm e} (s) \tilde{\eta}_{\rm e}(s') \rangle = D_{\beta} \frac{(s s')^{-\frac{\beta}{2}}}{s + s'} \quad \text{with} \quad \beta > 0 \label{noise} .
\end{align}
Here $\tilde{f}(s) = \int_{0}^{\infty} {\rm d} t e^{-s t} f(t)$ is the Laplace transform of the function $f(t)$.
In the time domain, the correlation function becomes ($t > t'$),
\begin{align}
\langle &\eta_{\rm e}(t) \eta_{\rm e}(t') \rangle = (-1)^{-\frac{\beta}{2}} \frac{D_{\beta}}{\Gamma^2\left(\frac{\beta}{2}\right)} (t-t')^{\beta-1} {\rm B} \left(- \frac{t'}{t-t'}, \frac{\beta}{2}, \frac{\beta}{2} \right), \label{noisecorr}
\end{align}
where ${\rm B}(x,a,b)$ is the incomplete Beta function \cite{abr64}. Equation \eqref{noisecorr} depends explicitly on time $t'$ and is therefore nonstationary.
For $\beta < 1$, it reduces to stationary fractional Gaussian noise, 
\begin{align}
\langle \eta_{\rm e}(t) \eta_{\rm e} (t') \rangle \simeq \frac{D_{\beta}}{\pi} \Gamma(1-\beta) \sin\left(\pi\frac{\beta}{2} \right) (t-t')^{\beta-1} \label{noisecorr_stat},
\end{align} 
in the long time limit, $t' \gg t-t'$, and is thus equivalent to Eq.~\eqref{noisecorr_old}.
In this regime, the external noise is akin to the internal one, but more strongly correlated for $\beta > \alpha$. 
By contrast, for $\beta > 1$, the external noise stays nonstationary and its variance grows with time as $\langle \eta_{\rm e}(t) \eta_{\rm e}(t) \rangle \sim t^{\beta-1}$. 
Figure~\ref{fig0} summarizes the properties of the noise correlations in the different regimes. 
We emphasize that our results will not depend on the specific form of the correlation function \eqref{noisecorr}, but only on its asymptotic behavior at long times. 

\begin{figure}[t]
\begin{center}
\includegraphics[width=0.8\textwidth, trim=20mm 5mm 25mm 15mm, clip]{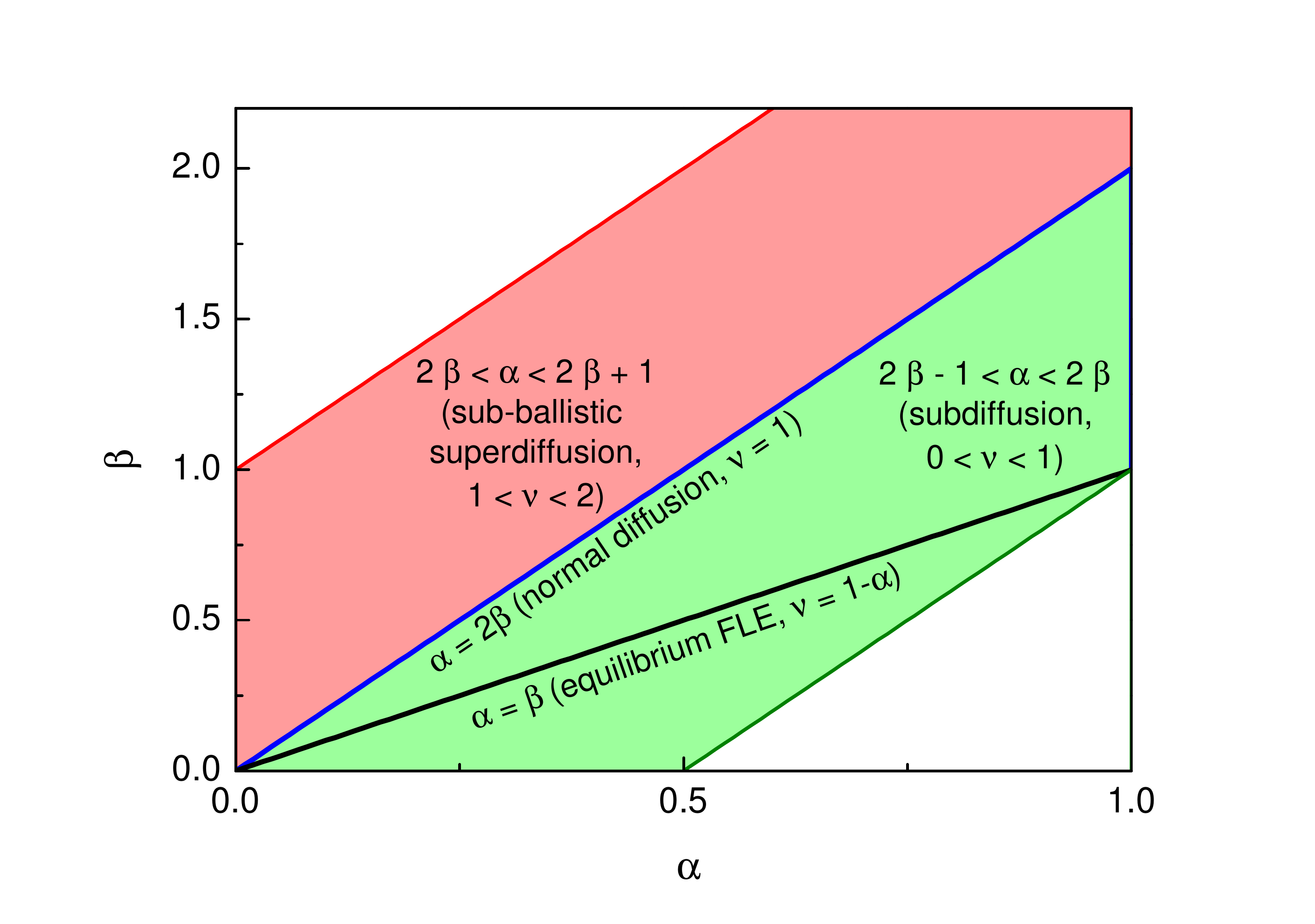}
\caption{The diffusive phases due to the external noise as a function of the exponents $\alpha$ of the memory kernel and $\beta$ of the external noise. The shaded area corresponds to the experimentally relevant regime $0 < \nu < 2$. \label{fig1}}
\end{center}
\end{figure}
\begin{figure}
\begin{center}
\includegraphics[width=0.8\textwidth, trim=20mm 10mm 25mm 15mm, clip]{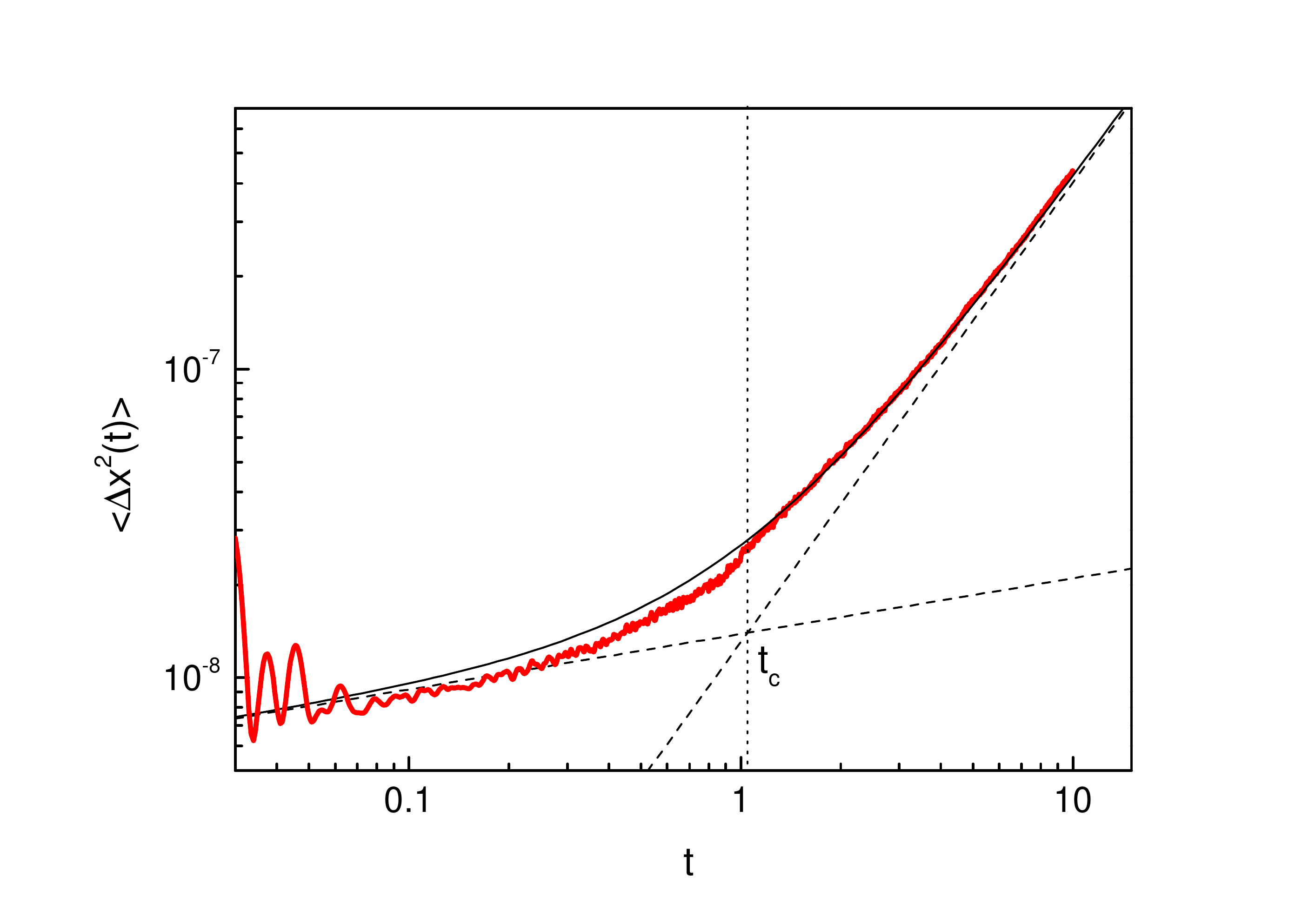}
\caption{Comparison between the asymptotic result for the mean-square displacement \eqref{MSDneq} (black) and the numerical simulations  of the Langevin equation \eqref{GLE} (red) for $\alpha = 0.82$ and $\beta = 2.13$. The transition from the subdiffusive equilibrium behavior \eqref{MSDneq} to superdiffusion is clearly visible and indicated by the dashed lines. The vertical dotted line marks the crossover time $t_c$, Eq.~\eqref{crossover}.\label{fig2} }
\end{center}
\end{figure}
To compute the contribution of the external noise force to the mean-square displacement \eqref{MSDeq}, we take the Laplace transform of Eq.~\eqref{GLE} (for $F_{\rm p} = 0$) and find,
\begin{align}
m (s^2 \tilde{x}(s) &- s x_0 - v_0) = - \tilde{\gamma}(s) (s \tilde{x}(s) - x_0) + \tilde{\eta}_{\rm i}(s) + \tilde{\eta}_{\rm e}(s) \label{GLElaplace},
\end{align}
where $x_0$, $v_0$ are the initial  position and velocity. 
Solving for $\tilde{x}(s)$, we obtain the Laplace transform of the position autocorrelation,
\begin{align}
\langle \tilde{x}(s) \tilde{x}(s') \rangle = \langle \tilde{x}(s) \tilde{x}(s') \rangle_{\rm i} + \tilde{g}(s) \tilde{g}(s') \langle \tilde{\eta}_{\rm e} (s) \tilde{\eta}_{\rm e}(s') \rangle, \label{corrlaplace}
\end{align}
where $\langle \tilde{x}(s) \tilde{x}(s') \rangle_{\rm i}$ denotes the part of the correlation function due to the internal noise (we have here assumed that  internal and external noise are statistically independent, since they have different sources). The kernel function $\tilde{g}(s)$ in Eq.~\eqref{corrlaplace} is defined as,
\begin{align}
\tilde{g}(s) = \frac{1}{s (m s + \tilde{\gamma}(s))} . \label{kernel}
\end{align}
The position autocorrelation in the time domain can be evaluated after Laplace inversion by  using the correlation function \eqref{noise}. We find,
\begin{align}
\langle x(t) x(t') \rangle &= \langle x(t) x(t') \rangle_{\rm i} + D_{\beta} \int_{0}^{{\rm min}(t,t')} dt'' F(|t-t'| + t'') F(t'') \label{xcorr} \\
\text{with} \quad &F(t) = \frac{1}{m}  t^{\frac{\beta}{2}+1} E_{\alpha+1,\frac{\beta}{2}+1} \left(-\frac{\gamma_{\alpha}}{m} t^{\alpha+1} \right), \nonumber
\end{align}
where $E_{a,b}(x)$ is the generalized Mittag-Leffler function \cite{bat55}.  
The asymptotic behavior of the mean-square displacement, for  $t = t' \gg (m/\gamma_{\alpha})^{1/(\alpha+1)}$, then follows as,
\begin{align}
\langle \Delta x^2(t) \rangle &\simeq \langle \Delta x^2(t) \rangle_{\rm i} + \frac{D_{\beta}}{\gamma_{\alpha}^2} \frac{1}{\nu \Gamma^2\left(\frac{\nu + 1}{2} \right)} t^{\nu} , \label{MSDneq}
\end{align}
with $\nu = \beta-2\alpha+1$. 
Here $\langle \Delta x^2(t) \rangle_{\rm i}$ is the equilibrium contribution to the mean-square displacement given by Eq.~\eqref{MSDeq} and the second term on the right stems from the external noise. 
Comparing the two expressions, we see that for $\beta > \alpha$, the contribution from the external noise dominates for long times. 
The latter  is subdiffusive for $\beta < 2\alpha$ and superdiffusive for $\beta > 2 \alpha$ (see Fig.~\ref{fig1}).
For the experimentally observed values of the diffusion exponent $0 < \nu < 2$ \cite{wei04,bur05,gol06,tre07,tre08,gal09,bru09}, we have $2\alpha - 1 < \beta < 2\alpha + 1$.
In Fig.~\ref{fig2}, the asymptotic result for the mean-square displacement Eq.~\eqref{MSDneq} is compared to those  of numerical  simulations of Eq.~\eqref{GLE}, showing excellent agreement.

To characterize the transition between the two diffusion regimes, we introduce the crossover time $t_c$, at which the contributions from the internal and external noise to the mean-square displacement \eqref{MSDneq} are of the same magnitude (see Fig.~\ref{fig2}). Using Eq.~\eqref{MSDeq}, we find, 
\begin{align}
t_c = \left( \frac{2 \nu \Gamma^2\left(\frac{\nu+1}{2}\right)}{\Gamma(2-\alpha)} \frac{k_B T \gamma_{\alpha}}{D_{\beta}} \right)^{\frac{1}{\nu+\alpha-1}} \label{crossover} .
\end{align}
The equilibrium mean-square displacement \eqref{MSDeq} correctly describes the dynamics  in the time window $(m/\gamma_{\alpha})^{1/(\alpha+1)} \ll t \ll t_c$, while the nonequilibrium external noise term dominates for $t\gg t_c$.

The behavior of the mean-square displacement \eqref{MSDneq} can be better understood by calculating  the velocity autocorrelation function, which  for long times  is given by ($t > t'$),
\begin{align}
\langle &v(t) v(t') \rangle \simeq \langle v(t) v(t') \rangle_{\rm i} + \frac{D_{\beta}}{\gamma_{\alpha}^2} \frac{1}{\pi} \Gamma(2-\nu) \sin\left(\pi\frac{\nu-1}{2}\right) (t-t')^{\nu-2} . \label{vcorr}
\end{align}
Expression \eqref{vcorr} is stationary and decays as a power law; it is negative for $\beta < 2 \alpha$ and positive for $\beta > 2 \alpha$. The latter correspond, respectively, to anticorrelated subdiffusion and correlated superdiffusion. The long-ranged nature of the velocity correlations is responsible for the anomalous dynamics of the system.

\section{Nonequilibrium Stokes-Einstein relation\label{SEC3}}

In contrast to the  equilibrium internal noise force \eqref{FDT}, the nonequilibrium external noise \eqref{noisecorr} may lead to a violation of the Stokes-Einstein relation \eqref{einstein}. Indeed, if the external noise is more strongly correlated than the thermal noise ($\beta > \alpha$), the asymptotic behavior of the mean-square displacement is given by Eq.~\eqref{MSDneq}, which is no longer related to the response \eqref{creepeq} via the temperature. 
However, we can establish a generalized Stokes-Einstein relation in the form, 
\begin{align}
\langle \Delta x^2(t) \rangle = 2 k_B T_{\rm eff}(t) J(t) , \label{einsteinneq}
\end{align}
by introducing  an effective time-dependent temperature,
\begin{align}
k_B T_{\rm eff}(t) &= \frac{D_{\beta}}{2 \gamma_{\alpha}} \frac{\Gamma(2-\alpha)}{\nu \Gamma^2\left(\frac{\nu + 1}{2}\right)} t^{\nu + \alpha - 1} \label{Teff}.
\end{align}
Effective temperatures are often  used to  measure  how far away from equilibrium a system operates \cite{cug97,cug11}. 
For $\beta > \alpha$, Eq.~\eqref{Teff} increases with time, indicating that the nonequilibrium properties of the system become more pronounced as time progresses. The latter can be understood by noting that the external noise injects energy into the system, driving it further away from equilibrium. 
For $\beta < \alpha$, the contribution from the external noise can be neglected asymptotically, and we have therefore $T_{\rm eff} = T$.

\begin{figure}[t]
\begin{center}
\includegraphics[width=0.8\textwidth, trim=20mm 10mm 25mm 15mm, clip]{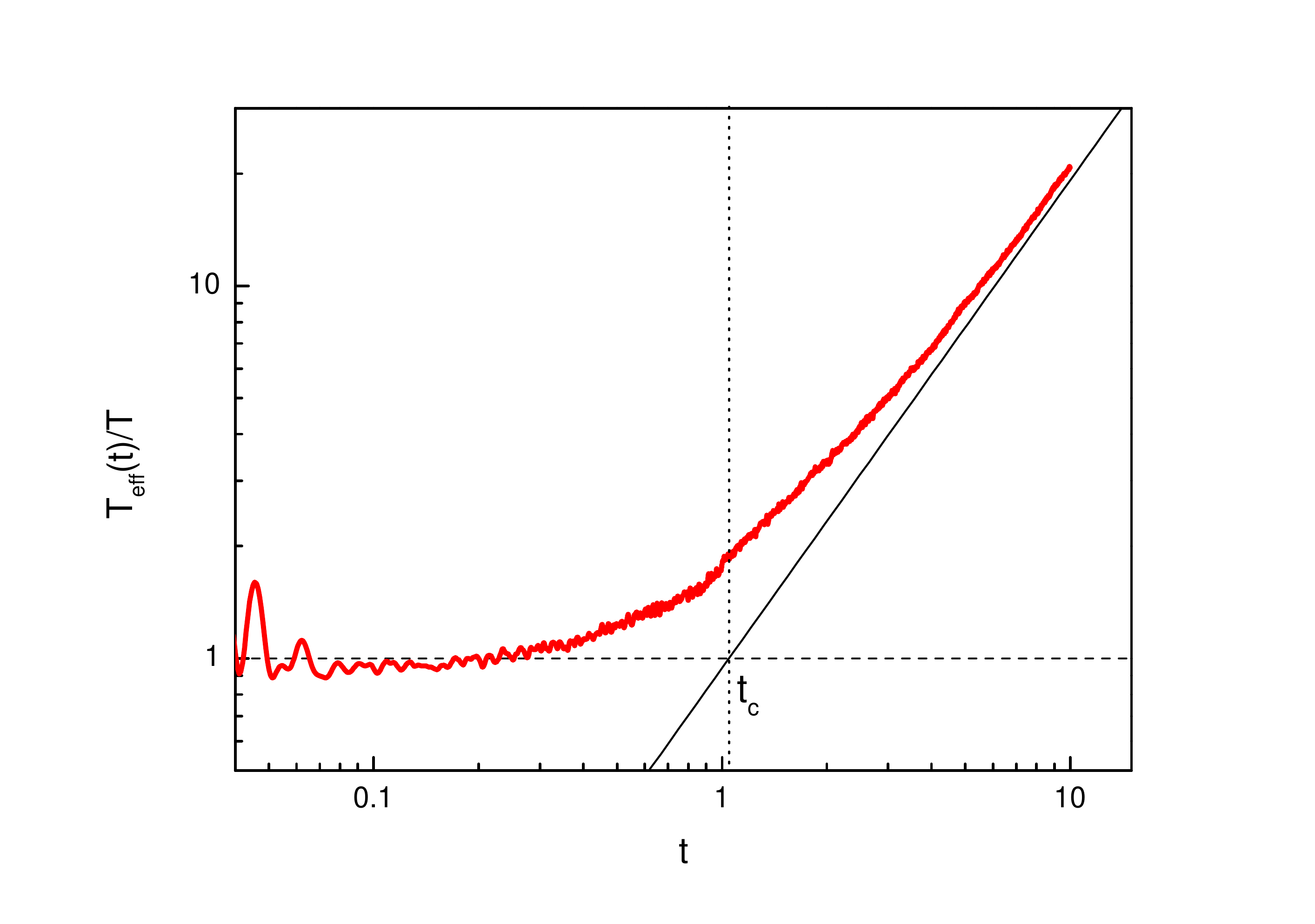}
\caption{Time-dependent effective temperature $T_{\rm eff}(t)/T$ for $\alpha = 0.82$ and $\beta = 2.13$ ($\nu = 1.49$). The red line is the result of numerical Langevin simulations, the solid black line corresponds to the asymptotic expression \eqref{Teff}. For short times, $t\ll t_c$, $T_{\rm eff}(t) \simeq T$. On the other hand, the effective temperature grows as a power law with exponent $\nu + \alpha - 1 \simeq 1.31$ for long times, $t\gg t_c$.\label{fig3}}
\end{center}
\end{figure}

The introduction of the effective temperature \eqref{Teff} might at first glance  appear purely formal. However, it gains physical significance by noting that it also measures the departure of the mean-square displacement \eqref{MSDneq} from its equilibrium expression \eqref{MSDeq}.  Indeed, for $t \ll t_c$, the effective temperature is equal to the actual physical temperature $T_{\rm eff}(t \ll t_c) = T$, since the equilibrium contribution from the thermal noise dominates at short times (see Fig.~\ref{fig3}). 
As a result, the asymptotic and transient behavior of the mean-square displacement are related via the effective temperature \eqref{Teff} as, 
\begin{align}
\langle \Delta x^2(t) \rangle = \frac{T_{\rm eff}(t)}{T} \langle \Delta x^2(t) \rangle_{t \ll t_c} \label{longshort} .
\end{align}
Equations \eqref{einsteinneq} and \eqref{longshort} provide two independent and complementary ways to determine the effective temperature: either by combining long-time active and passive measurements (as expressed by Eq.~\eqref{einsteinneq}), or by comparing short and long-time behavior in passive measurements alone (as indicated by Eq.~\eqref{longshort}). Combining both equations,  the transport properties of the system can thus be gained from passive measurements.  The generalized Stokes-Einstein relation \eqref{einsteinneq} with the effective temperature \eqref{Teff} therefore enable to connect passive and active microrheology in the nonequilibrium regime.

In practice, system properties are often measured in frequency space by exerting a periodic force on the tracer particle \cite{mas95,miz08,hof09}. It is thus desirable to define a frequency-dependent effective temperature.
If the probe particle is acted on by a periodic force, $F_{\rm p}(t) = F_0 \sin(\omega t)$, the response to this force may be characterized by the complex modulus $G(\omega)$, whose real and imaginary part describe the component of the force that is in-phase, respectively out-of-phase, with the response. From the expression \eqref{creepeq} for the creep function, we find,
\begin{align}
G'(\omega) &= \gamma_{\alpha} \omega^{1-\alpha} \sin\left(\pi \frac{\alpha}{2}\right) \nonumber, \\
G''(\omega) &= \gamma_{\alpha} \omega^{1-\alpha} \cos\left(\pi \frac{\alpha}{2}\right). \label{modulus}
\end{align}
As expected, the system behaves as a viscous liquid ($G'=0$) for $\alpha = 0$ and an elastic medium ($G''=0$) for $\alpha = 1$. 
The sublinear frequency dependence of the modulus for low frequencies has been experimentally observed in Refs.~\cite{wil08,hof09,mas07}.
For an equilibrium system, the imaginary part of the inverse complex modulus $G^{-1}(\omega)$ can be related to the power spectral density of the velocity $S_v(\omega)$ via,
\begin{align}
S_v(\omega) = - 2 k_B T \omega [G^{-1}]''(\omega) . \label{einsteinfreq}
\end{align}
The latter is an expression of the fluctuation-dissipation relation in frequency space \cite{rei65}.
As before,  this equilibrium   relation can be generalized to the nonequilibrium case with external noise by introducing an effective temperature \cite{pot04},
\begin{align}
S_v(\omega) &= - 2 k_B \mathcal{T}_{\rm eff}(\omega) \omega [G^{-1}]''(\omega) \label{einsteinfreq2}.
\end{align}
Since the velocity process is stationary in the considered range of parameters $\alpha$ and $\beta$, its spectral density is given by the Fourier transform of its stationary autocorrelation function according to the Wiener-Khinchine theorem \cite{rei65},
\begin{align}
S_v(\omega) = \int_{-\infty}^{\infty} {\rm d} \tau \, e^{-i \omega \tau} \langle v(t+\tau) v(t) \rangle = \frac{D_{\beta}}{\gamma_{\alpha}^2} \omega^{1-\nu} . \label{vPSD}
\end{align}
As a consequence, we can identify the frequency-dependent effective temperature by comparing  Eqs.~\eqref{modulus} and \eqref{einsteinfreq2}. We find,
\begin{align}
k_B \mathcal{T}_{\rm eff}(\omega) &= \frac{D_{\beta}}{2 \gamma_{\alpha}} \frac{1}{\cos\left(\pi \frac{\alpha}{2} \right)} \omega^{1-\nu-\alpha} \label{Teff2} .
\end{align}
The above definition of the effective temperature is  equivalent to the one introduced for the mobility in Ref.~\cite{pot04}. Two other effective temperatures have been defined in a similar way, via the the spectral density of the position in Ref.~\cite{jab08} and via the spectral density of the noise in  Ref.~\cite{gal09}. However,  the latter  are not suited to the superdiffusive, nonstationary regime as we will discuss below. 
The fact that $\mathcal{T}_{\rm eff}(\omega)$ increases with decreasing frequency for $\beta > \alpha$ indicates that the nonequilibrium properties of the system get more pronounced for low frequencies, corresponding to longer time scales, in complete analogy to the behavior of the effective temperature \eqref{Teff} in time; both  measure the distance of the system from equilibrium. 

In Eq.~\eqref{einsteinfreq2}  the viscous modulus was related to the power spectral density of the velocity. 
Does a similar relation hold for the power spectral density of the position? 
For the subdiffusive regime $\beta < 2\alpha$, the answer is yes; here, the two spectral densities are connected via,
\begin{align}
\omega^2 S_x(\omega) = S_v(\omega) \quad \text{for} \quad \beta < 2 \alpha \label{PSD} .
\end{align}
Even though the position process is nonstationary (its variance increases with time), its power spectral density does not depend on time in this regime. 
However, in the superdiffusive regime $\beta > 2 \alpha$,  the power spectral density now depends explicitly on time,
\begin{align}
S_x(\omega,t) \simeq \frac{D_{\beta}}{\gamma_{\alpha}^2 \nu \Gamma^2\left(\frac{\nu+1}{2}\right)} \omega^{-2} t^{\nu-1} \label{xPSD},
\end{align}
and cannot be related to the viscous modulus via a frequency-dependent temperature. 
Such a time-dependent spectral density is also found for the external noise \eqref{noise}:
\begin{align}
S_{\eta_{\rm e}}(\omega,t) \simeq D_{\beta} \left\lbrace \begin{array}{ll}
\omega^{-\beta} &\text{for} \; \beta < 2 \\[1 ex]
\frac{1}{(\beta-1) \Gamma^2\left(\frac{\beta}{2}\right)} \omega^{-2} t^{\beta-2} &\text{for} \; \beta > 2. 
\end{array} \label{noisePSD} \right.
\end{align}
The above unusual behavior is a direct consequence of the nonstationary properties of the external noise.
For values of the noise exponent $0 < \beta < 1$, the noise correlation function is stationary, $\langle \eta_{\rm e}(t) \eta_{\rm e}(t') \rangle \sim (t-t')^{\beta-1}$ (see Eq.~\eqref{noisecorr_stat}), and its exponent is simply related to the exponent of the power spectral density $S_{\eta_{\rm e}}(\omega) \sim \omega^{-\beta}$ via the Wiener-Khinchine theorem. 
However, for $1 < \beta < 2$,  even though the power spectral density is time independent, it would be wrong to conclude that the noise correlations also behave as $\langle \eta_{\rm e}(t) \eta_{\rm e}(t') \rangle \sim (t-t')^{\beta-1}$, since in this regime, the noise is actually nonstationary, and thus the Wiener-Khinchine relation no longer applies. 
Here, it is necessary to consider the full nonstationary noise correlation function \eqref{noisecorr}.

\section{Application to superdiffusion in living cells \label{SEC4}}

Let us now apply our formalism to the description of anomalous diffusion in biological cells.
The creep function and the mean-square displacement for muscle cells of adult mice were determined by means of active and passive microrheological measurements on the same sample in Ref.~\cite{gal09}. 
Sublinear growth of the creep function $J(t) \sim t^{0.18 \pm 0.02}$ was observed, which corresponds to $\alpha = 0.82 \pm 0.02$ in our model. 
For an equilibrium system without external noise, one would thus expect $\langle \Delta x^2(t) \rangle \sim t^{\nu}$ with $\nu = 0.18 \pm 0.02$ from the Stokes-Einstein relation \eqref{einstein}. 
This agrees well with the short-time behavior observed in the experiment, $\nu =0.12 \pm 0.01$ \cite{gal09}. 
For long times, however, the mean-square displacement was found to behave superdiffusively with an exponent $\nu =1.49 \pm 0.06$. According to Eq.~\eqref{MSDneq}, this corresponds to $\beta = 2.13 \pm 0.08$ from the active measurement of the creep function and $\beta = 2.25 \pm 0.07$ from the passive measurement of the short-time behavior of the mean square displacement. 
Both values exhibit good agreement within the error margins, confirming the validity of the generalized Stokes-Einstein relation \eqref{einsteinneq} connecting active and passive microrheology in the nonequilibrium regime.

The values of the diffusion exponent for short and long times appear to be universal for microrheological measurements performed on the cytoskeleton  of a large class of cells \cite{tre07,tre08}: the short-time subdiffusion exponent is typically around $0.2$, whereas the superdiffusion exponent is around $1.6$. 
This corresponds to a value of $\beta \simeq 2.2$ in good agreement with the values stated above. 
The resulting parameter $\alpha \simeq 0.8$ also agrees with measurements of the viscoelastic modulus \cite{hof09,mas07}.
The crossover time $t_c$ is found to be on the order of $1 s$ \cite{gal09,tre08}, and is therefore easily observed. %Together with the values for the exponents $\nu$ and $\alpha$, this allows to estimate the ratio $D_{\beta}/(k_B T \gamma_{\alpha}) \simeq 2.66 s^{1.31}$, which measures the relative strength of the external and internal noise.
The results of our analysis shows that the external noise is nonstationary for  $\beta > 2$ and  that the noise  power spectral density \eqref{noisePSD} is thus explicitly time-dependent, $S_{\eta}(\omega,t) \sim \omega^{-2} t^{\beta-2}$. The latter is  in agreement with the $\omega^{-2}$ behavior of the intracellular stress fluctuations measured in Ref.~\cite{lau03} for a fixed time, and found in Ref.~\cite{mac08} for a model of the active cytoskeletal actin network. We emphasize that these findings cannot be reproduced using the stationary noise correlation \eqref{noisecorr_old} of Ref.~\cite{bru09}. In this experiment, $\beta\simeq 0.42$, a value corresponding to the stationary regime. The latter is due to a relatively large subdiffusion exponent $\nu$ close to 1, a value also observed in a couple of other experiments \cite{wei04,tse02}. 

\begin{figure}[t]
\begin{center}
\includegraphics[width=0.8\textwidth, trim=20mm 12mm 25mm 15mm, clip]{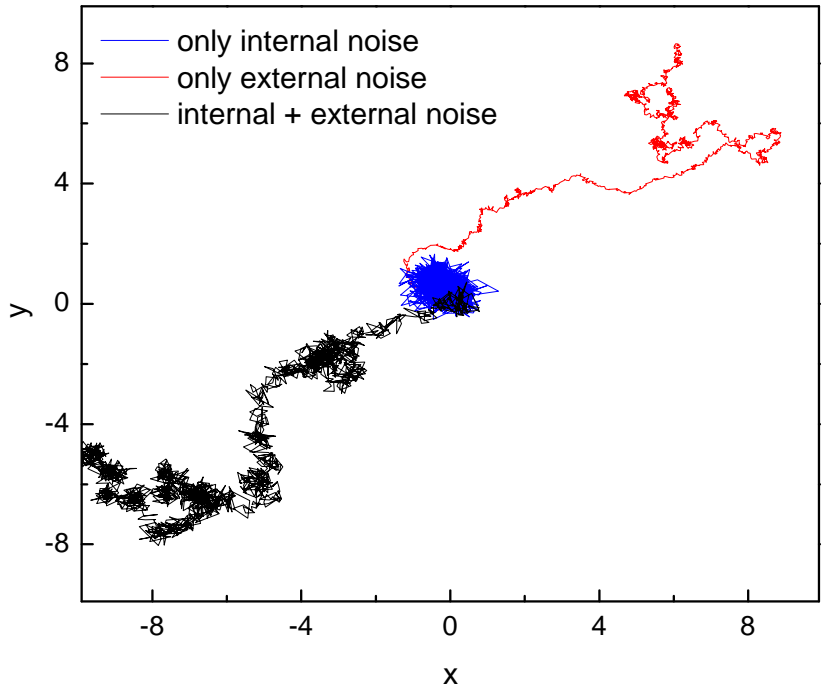}
\caption{Typical trajectories obtained by simulating the fractional Langevin equation \eqref{GLE} for $\alpha = 0.8$, $\beta = 2.2$. The blue trajectory (center) corresponds to internal noise, the red one (top right) to external noise only. The black one (bottom left) is obtained for the  combination of both noises. In the latter, the particle   covers a large distance and, at the same time, explores its immediate surroundings. All trajectories start at the origin and have the same duration. \label{fig4}}
\end{center}
\end{figure}

\section{Discussion \label{SEC5}}

Our model provides an extension of the fractional Langevin equation  to describe  anomalous diffusion in cells in the presence of nonequilibrium external noise. 
We have used the fact that the transient behavior in  passive measurements is  determined by the equilibrium properties to   restore the connection between active and passive microrheology via a measurable effective temperature. We have confirmed the validity of the generalized Stokes-Einstein relation \eqref{einsteinneq} by direct comparison with  the  experimental data of Ref.~\cite{gal09}. The observation that the response of living cells appears generic  for a large class of cells \cite{tre07,tre08} hints at the widespread applicability  of our formalism.

Typical trajectories obtained from numerical simulations of the Langevin equation are shown in Fig.~\ref{fig4}. They exhibit remarkable similarity to experimentally observed trajectories \cite{gal09,bur05}. 
Under the sole action of the internal thermal noise, the particle thoroughly explores its immediate surroundings (blue trajectory). 
It has been shown that this subdiffusive motion is advantageous for reactions in the cell, leading to an increased probability of finding a nearby target site \cite{gui08}. By contrast, in the presence of both internal (equilibrium) and external (nonequilibrium) noise, the dynamics becomes superdiffusive (black trajectory), and the particle  covers large distances while still exploring the surrounding area. The latter  allows for both fast transport and facilitated reactions---this last aspect is missing when only the nonequilibrium noise is present (red trajectory).

\pagebreak

The origin of superdiffusion in living cells is thought to be the result of the collective behavior of molecular motors, switching on and off randomly, giving rise to correlated noisy motion \cite{lau03,wil08,bur05}. A value of   $\beta > 2$  indicates that the external noise becomes slightly more correlated over time, as seen in Fig.~\ref{fig0}. Our analysis thus  suggests increasing cooperative action on part of the individual molecular motors in the nonstationary regime. An experimental confirmation of the resulting time-dependence of the power spectrum would therefore uncover some interesting new aspects of cell dynamics.

{\bf Acknowledgments:} This work was supported by the Focus Area Nanoscale of the FU Berlin and the DFG (Contract No 1382/4-1). AD also thanks the Elsa-Neumann graduate funding for support.

%% PNAS does not support submission of supporting .tex files such as BibTeX.
%% Instead all references must be included in the article .tex document. 
%% If you currently use BibTeX, your bibliography is formed because the 
%% command \verb+\bibliography{}+ brings the <filename>.bbl file into your
%% .tex document. To conform to PNAS requirements, copy the reference listings
%% from your .bbl file and add them to the article .tex file, using the
%% bibliography environment described above.  

%%  Contact pnas@nas.edu if you need assistance with your
%%  bibliography.

% Sample bibliography item in PNAS format:
%% \bibitem{in-text reference} comma-separated author names up to 5,
%% for more than 5 authors use first author last name et al. (year published)
%% article title  {\it Journal Name} volume #: start page-end page.
%% ie,
% \bibitem{Neuhaus} Neuhaus J-M, Sitcher L, Meins F, Jr, Boller T (1991) 
% A short C-terminal sequence is necessary and sufficient for the
% targeting of chitinases to the plant vacuole. 
% {\it Proc Natl Acad Sci USA} 88:10362-10366.

%% Enter the largest bibliography number in the facing curly brackets
%% following \begin{thebibliography}

\end{document}